\documentclass[pra,twocolumn,aps,amsmath,showkeys,showpacs,floatfix,footinbib,
reprint]
{ revtex4-1}
\usepackage{mathptmx}
\usepackage{dcolumn}
\usepackage{amsmath,dsfont,amsfonts}
\usepackage{bm}
\usepackage{color}
\usepackage{latexsym}
\usepackage[english]{babel}
\usepackage{times}
\usepackage{psfrag,graphicx}
\usepackage{epsf}
\usepackage{natbib}
\usepackage{epstopdf}
\usepackage{appendix}
\usepackage[colorlinks=true,citecolor=blue,linkcolor=blue]{hyperref}
\usepackage{ulem} 
 \newcommand{\ket}[1]{\left|#1\right\rangle} 
 \newcommand{\bra}[1]{\left\langle#1\right|} 

 \def\vec#1{{\boldsymbol{#1}}}  

\begin{document}


\title{Quantum magnetism of ultracold atoms with a dynamical pseudospin degree of freedom}

\author{ Tobias Gra\ss$^1$, Alessio Celi$^1$, and Maciej Lewenstein$^{1,2}$}

\affiliation{$^1$ICFO-Institut de Ci\`encies Fot\`oniques, Parc Mediterrani
de la Tecnologia, 08860 Barcelona, Spain}
\affiliation{$^2$ICREA-Instituci\'o Catalana de Recerca i Estudis Avan\c cats, 
08010 Barcelona, Spain}

\begin{abstract}
We consider bosons in a Hubbard lattice with an SU($\cal N$) pseudospin degree
of freedom which is made dynamical via a coherent transfer term. It is shown
that, in the basis which diagonalizes the pseudospin coupling, a generic hopping
process affects the spin state, similar to a spin-orbit coupling. This results,
for the system in the Mott phase, in a ferromagnetic phase with variable
quantization axis. In extreme cases, it can even give rise to antiferromagnetic
order.
\end{abstract}

\maketitle

\section{Introduction}
Ultracold atoms in optical lattices are almost ideal realizations of different
Hubbard models. In certain limits, these models can be directly mapped on spin
models which are the key for understanding quantum magnetism and related
phenonema like antiferromagnetism or spin liquids \cite{auerbach94,mlbook}. A
particularly rich behavior can be explored by filling the lattice with
multi-component atoms in Mott states. The most prominent example, recently
realized experimentally \cite{greif12}, is the two-component Fermi gas. In the
Mott phase with one atom per site,  such system is perfectly described by the
antiferromagnetic Heisenberg model \cite{hof02}. The use of fermionic alkali
earth atoms in optical lattices has been proposed to study SU($\cal N$)
magnetism for $\cal N$ much larger than 2 \cite{gorshkov-sun}. Attention has
also been put on bosonic two-component systems \cite{altman-njp}, or bosonic
spinor gases with $F=1$ \cite{demler-zhou,imambekov,yip,garcia-ripoll04} or
$F=2$ \cite{barnett06}. 

One can further enrich such systems by a laser coupling of the atomic states. If
spatially dependent, such a coupling connects internal with external degrees of
freedom, and can thus be interpreted a non-Abelian artificial gauge field
\cite{dalibard}. The Mott transition in a bosonic Hubbard model is dramatically
modified by the presence of such a field \cite{indianpra}, and deep in the Mott
phase the gauge field supports phases with exotic magnetic ordering
\cite{trivediBH,galitski12}. Recently, it has been pointed out in Ref. \cite{extradim} 
that the internal degrees of freedom can also be used for simulating  an
``extradimension'', once the 
internal states are properly coupled to provide the hopping between the synthetic sites given by the atomic species. 

In this paper, we study the case of a spatially homogeneous coupling of the
atomic states, and analyze the Mott phases in an $\cal N$-component Bose-Hubbard
model in the presence of coherent transfer between the internal states.  In the
extradimension picture of Ref. \cite{extradim}, the internal degree of freedom
becomes equivalent to a compactified spatial dimension. Assuming SU($\cal N$)
symmetry, we show in Sec. \ref{system} that, in the appropriate spin basis, the
internal hopping acts as an external magnetic field. It is responsible for a
linear Zeeman shift lifting the degeneracy between the internal states. 

In Sec. \ref{sect:sun}, we then focus on a scenario where SU($\cal N$)
symmetry is broken, and consider systems with state-dependent hopping
strengths, as in the case of spin-dependent lattices \cite{liu04}. 
Quite generally, a hopping process will then also change a particle's internal
state in the eigenbasis of the coupling. As it has been proposed in Ref.
\cite{andre-shakes}, by shaking the optical lattice it is even possible to
reverse the sign of the hopping term. In combination with spin-dependent
lattices, this technique becomes species-selective and allows for generating
non-Abelian hopping terms \cite{hauke-shakes}. This includes the case of a
hopping of the form $J \sigma_z$, with $\sigma_z$ a Pauli matrix in SU(2), and
similar expressions for higher spin. We extend our study to such extreme
deviations from the SU($\cal N$)-symmetric hopping, and carefully analyze the
SU(2) scenario. We find that deviations from an SU($\cal N$)-symmetric
hopping rotate the quantization axis of a ferromagnetic phase. The full reversal
of one hopping strength gives rise to a spin-rotated superexchange interaction,
which favors unmagnetized states. This allows for a transition to an
antiferromagnetic or checkerboard phase, that is to say, the superposition of
pseudo-spin states becomes position-dependent following a crystal structure. 
In the extradimension picture, in which the different pseudo-spin states become
different sites, such structures become density structures.

Afterwards, in Sec. \ref{model}, we consider an SU($\cal N$) symmetry breaking
in the interaction term.
In particular, we assume the interspecies density-density interaction as a free parameter.  
The resulting model interpolates between an SU(${\cal N}$) Bose-Hubbard system
in $d$ dimensions, and ${\cal N}$ copies of a Bose-Hubbard system in $d+1$ space
dimensions, respectively.
Such model displays a rich Mott regime which can be perturbed to give rise to different phases.
In particular, we focus on the parameter region that admits as
degenerate
ground states Mott configurations with the number $n$ of particles per
site being non-commensurable 
with the number of species ${\cal N}$. We study in detail the paradigmatic example of $n= q {\cal N} +1$, with $q$ integer, i.e. 
one spin component is occupied by $q+1$ particles per site, while the other components are occupied with $q$ ones. 
At the perturbative level, the hopping terms induce a novel Potts-like effective
Hamiltonian that displays different quantum phases.
The different phases can be detected in time-of-flight absorption pictures by
applying real magnetic fields for a Stern-Gerlach-type measurement.

\section{System \label{system}}

We consider an $\cal N$-component Bose gas in a hypercubic optical lattice in
$d$
dimensions. The physics is well described by a Bose-Hubbard (BH)
Hamiltonian $H=H_I+H_J+H_0$, where $H_0$ is the local interaction term, $H_J$
the hopping term, and $H_I$ the coherent transfer (internal hopping) between the
pseudospin components. For $H_0$ we write:
\begin{align}
 H_0 = \sum_{i,\sigma} \left[\left( \frac{U}{2} \hat n_i^{\sigma}(\hat n_i^{\sigma}-1)
- \mu \hat n_i^{\sigma} \right) +
\frac{U_d}{2}\sum_{\sigma'}\hat n_i^{\sigma}\hat n_i^{\sigma'}\right],
\end{align}
with $\hat n_i^{\sigma} \equiv \hat a_i^{(\sigma)\dagger} \hat
a_i^{(\sigma)}.$ The operator $\hat a_i^{(\sigma)\dagger}$ creates a particle on
site $i$ in the pseudospin state labeled by $\sigma$. The parameters $U$ and
$U_d$ fix the (possibly) spin-dependent interaction strength. Many atoms,
amongst them $^{87}$Rb, possess hyperfine states with almost the same $s$-wave
scattering lengths, thus, they are approximately described by $H_0$, with
$U\approx U_d$. 
is $U=U_d$. 
The chemical potential $\mu$ allows to fix the total number of
particles per site.

Note that $H_0$ is quadratic in $\hat n^\sigma$, $H_0= \frac 12 {\bf v}^{\rm t}\cdot M \cdot {\bf v} -{\bf w}\cdot {\bf v}$, where 
${\bf v}^{\rm t}\equiv (\hat n^1,\dots,\hat n^{\cal N})$, 
and can be easily minimized 
by diagonalizing $M_{\sigma\rho}\equiv u+(1-u)\delta_{\sigma\rho}$, 
with  $u\equiv \frac {U_d}U$, $\sigma,\rho=1,\dots{\cal N}$, and  ${\rm w}_\sigma= \frac {\mu-U/2}U$. 
The content and the dimension of the minimal energy subspace depend strongly on the values of $u$ and  of the chemical potential $\mu$, see Appendix \ref{mott} for details.

For the external hopping, $H_J$, we take into account a possibly spin-dependent
nearest-neighbor tunneling:
\begin{align}
 H_J &= -\sum_{\sigma} J^{(\sigma)} \sum_{\langle ij\rangle} \hat
a_i^{(\sigma)\dagger} \hat a_j^{(\sigma)}.
\end{align}
Here, $J^{(\sigma)}$ is the pseudospin-dependent tunneling strength. 

A coherent transfer term $H_I$ locally replaces a $\sigma$ particle by a
$\sigma\pm 1$ particle. For convenience, we choose periodic boundaries for this
``internal'' hopping, that is, we shall take the value of $\sigma$ modulo
${\cal N}$:
\begin{align}
\label{HI}
 H_I = - I \sum_{i,\sigma} \left(\hat a_i^{(\sigma)\dagger} \hat
a_i^{({\rm mod} [\sigma+1,{\cal N}])} e^{i \phi} + \mathrm{H.c.} \right),
\end{align}
Experimentally, this term can be implemented by a resonant radio frequency in the linear Zeeman splitting regime 
of the hyperfine states (for open boundaries conditions) or by Raman lasers (for periodic boundary conditions, 
 in the quadratic Zeeman splitting regime for ${\cal N}>3$) 
shining onto the atoms, see \cite{Celi13}.
The laser intensity defines the coupling strength $I>0$, and the
photons may also imprint a phase angle $\phi$. Note that for ${\cal N}\phi\neq
2\pi {\cal Z}$, under a full loop in the species space the state acquires
a non-trivial phase. In the extradimension picture, this is equivalent to flux compactification of the synthetic $d+1$ dimension on a circle, with a magnetic flux piercing it.   
 The Hamiltonian $H_I$ can be expressed as a circulant matrix $C$, such that $H_I=I C$ with
\begin{align}
 C=-
\begin{bmatrix}
c_1     & c_{2} & \dots  & c_{{\cal N}-1} & c_{{\cal N}}  \\
c_{{\cal N}} & c_1    & c_{2} &         & c_{{\cal N}-1}  \\
\vdots  & c_{{\cal N}}& c_1    & \ddots  & \vdots   \\
c_{3}  &        & \ddots & \ddots  & c_{2}   \\
c_{2}  & c_{4} & \dots  & c_{{\cal N}} & c_1 \\
\end{bmatrix},
\end{align}
with $c_2=e^{i\phi}$ and $c_{{\cal N}}=e^{-i\phi}$, while $c_i=0$ for $i=1$ and
$3 \leq i \leq {\cal N}-1$.
Eigenvalues $\lambda_{\alpha}$ and the corresponding eigenvectors $\vec{v}_{\alpha}$ of this matrix are given in terms of exponential terms $\omega_{\alpha} \equiv \exp[(2\pi i/N)\alpha]$:
\begin{align}
 \vec{v}_{\alpha} = \frac{1}{\sqrt{{\cal N}}}(\omega_{\alpha}^0,
\omega_{\alpha}^1, \cdots, \omega_{\alpha}^{{\cal N}-1})^T,
\end{align}
and
\begin{align}
\label{lambda}
 \lambda_{\alpha} = 2\cos[\phi+(2\pi i/{\cal N})\alpha].
\end{align}
We note that the eigenvectors are independent from the phase shift $\phi$. On
the other hand, the eigenvalues can be tuned by this parameter. In particular,
we can make any $\vec{v}_{\alpha}$ the ground state of $H_I$. Alternatively, any
pair $\vec{v}_{\alpha}$ and $\vec{v}_{{\rm mod}[\alpha+1,{\cal N}]}$ can be made
a two-fold degenerate ground state. 
With the choice of $H_I$ as in Eq.~(\ref{HI}),
higher degeneracies are not
possible, but it is worth to notice that by turning on next-to-nearest
neighbor coupling of the species as well, 
fourfold degeneracy of the eigenvalues can be achieved. Such coupling can be realistically laser induced as the ones described above.  

It is convenient to introduce a matrix $V=[\vec{v}_1, \cdots, \vec{v}_{\cal N}]$
which
transforms from the original pseudospin basis into the basis of eigenstates of
$H_I$. Starting by the Fock operators in the old basis organized as an 
SU$({\cal N})$-vector operator $\vec{a}_i \equiv (\hat a_i^{(1)}, \cdots,
\hat a_i^{({\cal N})})^T$, which in every component annihilates a particle in the
original basis, we obtain the corresponding  Fock operators in the novel basis as  $\vec{A}_i \equiv V
\vec{a}_i\equiv (\hat A_i^{(1)}, \cdots, \hat A_i^{({\cal N})})$, which
componentwise
annihilates a particle in the state $\vec{v}_{\alpha}$.

Let us first discuss the case where $H_0$ and $H_J$ are SU($\cal N$) symmetric,
that
is $J^{(\sigma)}=J$ and $U=U_d$. We can write the full Hamiltonian as
\begin{align}
\label{HSUN}
 H= & \sum_i \left( \frac{U}{2} \hat N_i(\hat N_i-1) - \mu \hat N_i \right) -J
\sum_{\langle ij\rangle} \vec{A}_i^{\dagger} \vec{A}_j +
\nonumber \\ & I \sum_{i,\alpha}
\lambda_{\alpha} \hat A_i^{(\alpha)\dagger}A_i^{(\alpha)}.
\end{align}
We can associate the eigenvalues $\lambda_{\alpha}$ with a magnetic quantum
number. The internal coupling $I$ then can be interpreted a Zeeman
shift
$I\lambda_{\alpha}$ experienced by the state $\alpha$. This analogy to the
atomic finestructure is best drawn if the levels are equally spaced. 
For $H_I$ of the form \eqref{HI}, this condition can be realized by choosing
the proper $\phi$ for ${\cal N}\leq4$, but not for higher values of ${\cal
N}$
(at least if we limit to nearest neighbor species coupling in $H_I$).
In fact, in order to get an equally spaced spectrum  the coupling $I$ should depend on $\sigma$.
For real nearest neighbor hoppings, $H_I$ has to be taken proportional to the normalization of the raising operator $F_+= F_x+i F_y$ for fixed  
total angular momentum $F$ such that $I^{(\sigma)}= I
\sqrt{({\cal N}-\sigma)\sigma }$ for ${\cal N}=2F+1$.  
Note that this is exactly the coupling induced by Raman lasers in the far-detuned regime, as shown in \cite{Goldman13,Dudarev04,Juzeliunas13,Hugel13} and considered in \cite{Celi13}  
to simulate synthetic edge states with a synthetic dimension.

It is obvious that the last term of Eq. (\ref{HSUN}) breaks the SU($\cal N$)
symmetry. The ground state properties of the system then become independent from
the existence of the ``internal dimension'', that is, the extradimension
vanishes. For special choices of $\phi$, however, a two-fold degeneracy may
remain.

\section{Magnetic orderings in systems with SU($\cal N$) symmetry-breaking
hopping
terms}\label{sect:sun}

Deviations from SU($\cal N$) symmetry in the hopping naturally occur if the 
atomic states possess different polarizabilities. One may also generate extreme
deviations artificially using techniques for manipulating the hopping term. Such
techniques have been developed in the context of simulating gauge fields. As
discussed in Ref. \cite{hauke-shakes} for SU(2) systems, it is, for instance,
possible to reverse the hopping in one
component via shaking.

How deviations from SU($\cal N$) symmetry enrich the physics of the system
becomes
transparent when we transform $H_J$ into the $\vec{v}_{\alpha}$ basis which
diagonalizes $H_I$. Defining a vector $\vec{J} \equiv (J_1, \cdots, J_N)^T$ with
the different hopping parameters in its components, we
construct a matrix
\begin{align}
 Q_{kl} \equiv \left\{ \begin{array}{cl} \vec{J} \cdot \vec{v}_{k-l}, & \mbox{if
} k \geq l \\ \vec{J} \cdot \vec{v}_{N+k-l}, & \mbox{else} \end{array}\right.
\end{align}
With this matrix, we can write the hopping term as
\begin{align}
 H_J = \sum_{\langle ij\rangle}
 \vec{A}_i^{\dagger}\cdot Q \cdot\vec{A}_j.
\end{align}

Thus, in the new basis, hopping processes in $H_J$ in general not only change the external position of the particle, but also its internal state. 
The Hamiltonian is thus equivalent to one coupled to some constant non-Abelian gauge field. Hopping processes $\alpha \rightarrow \alpha$, 
that is, those which do not change the internal state have a hopping strength $j_N \equiv \vec{J}\cdot\vec{v}_N=\frac{1}{N}\sum_{\sigma=1}^N J^{(\sigma)}$. 
Hopping processes $\alpha \rightarrow {\rm mod}[\alpha+k,N]$ have a hopping strength $j_{k} \equiv \vec{J}\cdot\vec{v}_{k}$. 
Hence, the hopping term is generically complex. Note that $j_{k} = j_{N-k}^*$, that is, they can not independently be chosen.

Setting $J^{(\sigma)} \ll U$, we are deep in the Mott phase, and the Hamiltonian is solved by a Fock state of $n$ atoms per site. 
The number $n$ is tuned by the chemical potential $\mu$, which in the following is chosen such that $n=1$.

In the remainder of this section, we will first discuss in detail the
consequences of such symmetry-breaking hopping on the Mott phase of a
two-component Bose system. Afterwards, we will also take a brief look on
systems with ${\cal N}>2$ components.

\subsection{Two-component system}

In SU(2), the hopping matrix $H_I$ is explicitly diagonalized by the operators 
\begin{align}
 \hat A_i & \equiv \frac{1}{\sqrt{2}} \left( \hat{a}_i^{(a)} +
e^{i\phi} \hat{a}_i^{(b)} \right), \\
 \hat B_i & \equiv \frac{1}{\sqrt{2}} \left( \hat{a}_i^{(a)} -
e^{i\phi} \hat{a}_i^{(b)} \right).
\end{align}
In this basis, it reads
\begin{align}
 H_I = - I \sum_i \left( \hat A_i^{\dagger} \hat A_i - \hat B_i^{\dagger} \hat
B_i \right).
\end{align}
This expression shows that the phase $\phi$ is completely absorbed in the
operators $\hat A$ and $\hat B$. The ground state now is uniquely given by the
state $\ket{A}$.

Transforming $H_J$ into the $\hat A,\hat B$ basis which
diagonalizes the local problem, we get:
\begin{align}
H_J &=
\sum_{\langle ij\rangle} \Big[ -J_+ \left( \hat A_i^{(\sigma)\dagger} \hat
A_j^{(\sigma)} + \hat B_i^{(\sigma)\dagger} \hat
B_j^{(\sigma)} +\mathrm{H.c.} \right) 
- \nonumber \\ & 
J_- \left( \hat A_i^{(\sigma)\dagger} \hat
B_j^{(\sigma)} + \hat B_i^{(\sigma)\dagger} \hat
A_j^{(\sigma)} +\mathrm{H.c.} \right)\Big],
\end{align}
with $J_+ \equiv (J^{(a)}+J^{(b)})/2$ and $J_- \equiv (J^{(a)}-J^{(b)})/2$.
While in the SU(2)-symmetric case, the off-diagonal terms vanish, the opposite
occurs if we reverse one hopping strength, $J^{(a)}=-J^{(b)}$.

The lowest-order contribution to the effective Hamiltonian is quadratic in $H_J$, and connects any site $i$
with its nearest neighbors $j$. We denote the low-energy states on such an $ij$
pair by $\ket{AA}$, $\ket{BB}$, $\ket{AB}$, $\ket{BA}$. We obtain the effective
Hamiltonian $H_{\rm eff} = H_{I} + \sum_{\langle ij \rangle} H_{\rm
eff}^{ij}$, where  
\begin{widetext}
\begin{align}
\label{Heff}
- H_{\rm eff}^{ij} = &
\ket{AA}\bra{AA} \left( \frac{4 J_+^2}{U} + \frac{2
J_-^2}{U+2I} \right) + 
\ket{BB}\bra{BB} \left( \frac{4 J_+^2}{U} + \frac{2
J_-^2}{U-2I} \right)+
\nonumber \\ &
\left(\ket{AA}\bra{AB}+\ket{BB}\bra{AB}+\ket{AA}\bra{BA}+\ket{BB}\bra{BA} + {\rm
H.c.}\right)  \left( \frac{2 J_+ J_-}{U-I} + \frac{2
J_-J_+}{U+I} \right)+
\nonumber \\ &
(\ket{AB}\bra{AB}+\ket{BA}\bra{BA}) \left( \frac{4 J_+^2}{U} + \frac{2
J_-^2}{U-2I} + \frac{2
J_-^2}{U+2I} \right)+
\nonumber \\ &
(\ket{AA}\bra{BB} + {\rm H.c.})\frac{2J_-^2}{U}+
(\ket{AB}\bra{BA} + {\rm H.c.})\frac{2J_+^2}{U}.
\end{align}

Within an SU(2) spin notation, the effective Hamiltonian reads
\begin{align}
\label{spinH}
 H_{\rm eff} =& \sum_{\langle ij \rangle}
\frac{4J_+^2}{U} \vec{S}_i\cdot \vec{S}_j + \frac{4J_-^2}{U} \left(S_i^x S_j^x
-
S_i^y S_j^y \right) - 2J_-^2
\left(\frac{1}{U+2I}+\frac{1}{U-2I}\right)S_i^zS_j^z +
\nonumber \\ &
d \sum_i S_i^z \left[-\frac{1}{d}2I + 2 J_-^2
\left(\frac{1}{U+2I}-\frac{1}{U-2I} \right)\right] + S_i^x \left[8
J_+J_- \left(\frac{1}{U+I}+\frac{1}{U-I} \right) \right].
\end{align}
In the second line, the term stemming from $H_J$ is enhanced against the local
term $H_I$ by a factor $d$ counting the number of spatial dimensions.
\end{widetext}

\begin {figure*}
\hspace{-3cm}\includegraphics[width=0.34\textwidth, angle=-90]{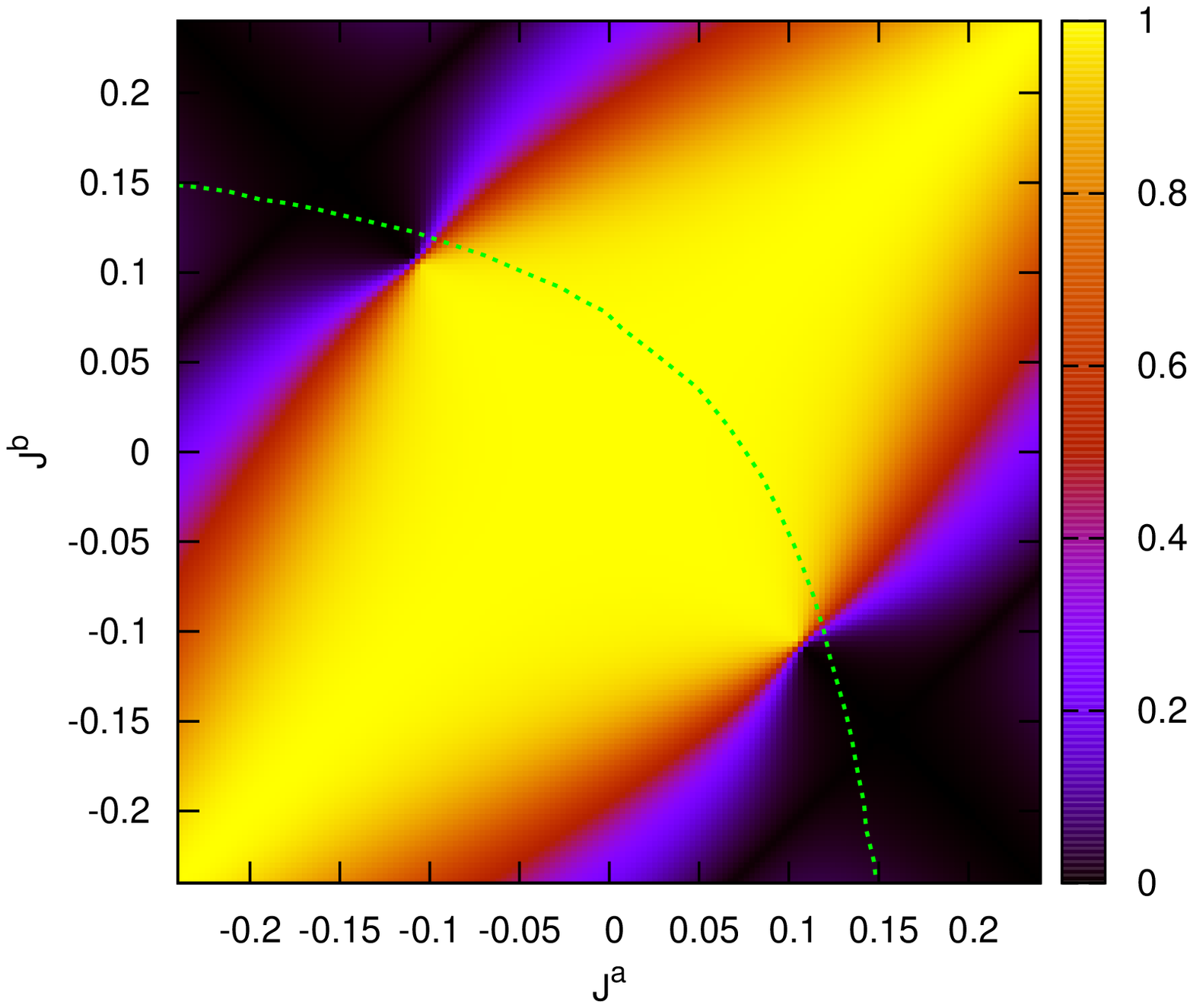}
\hspace{-3cm}\includegraphics[width=0.33\textwidth, angle=-90]{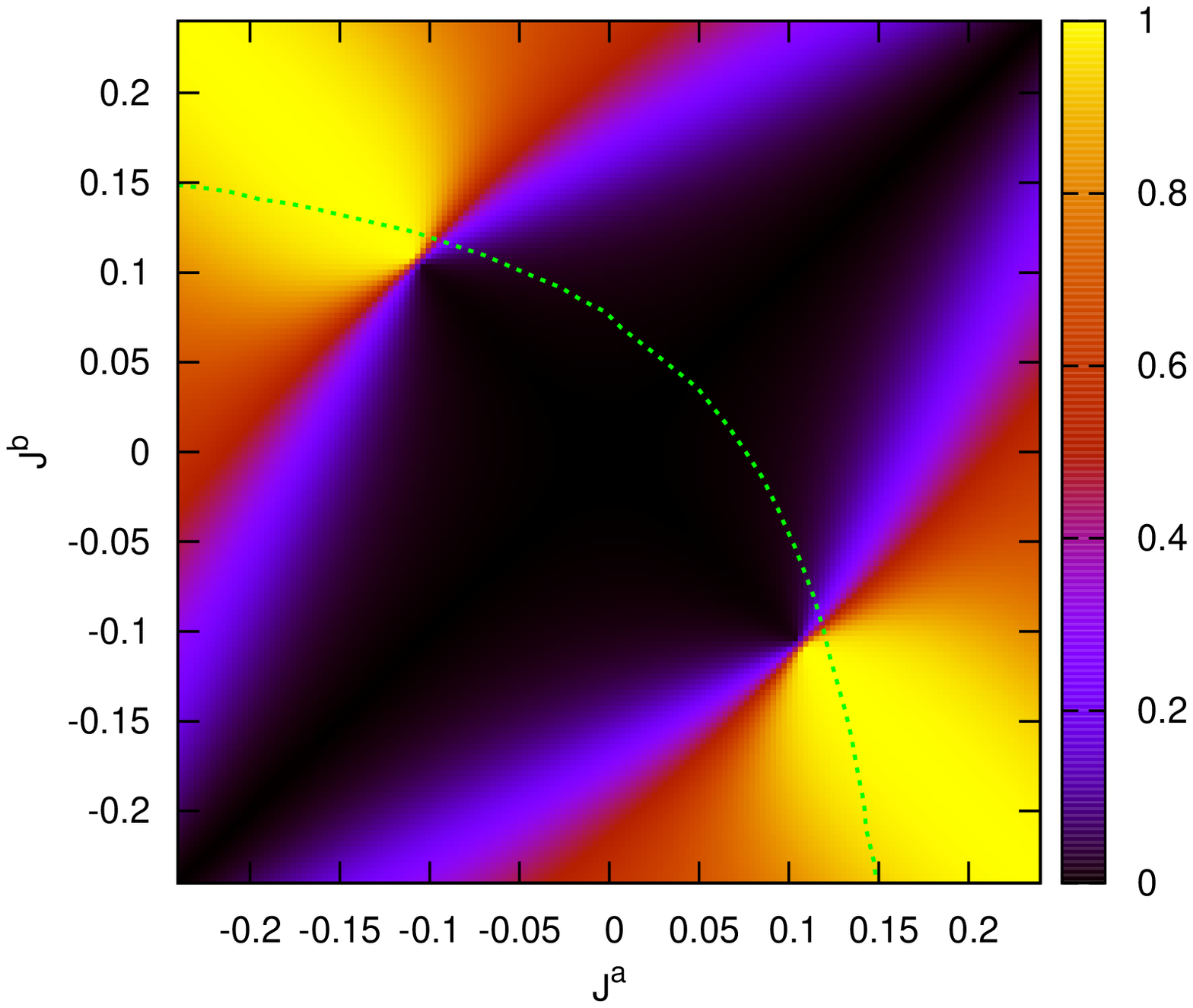}
\hspace{-3cm}\includegraphics[width=0.33\textwidth, angle=-90]{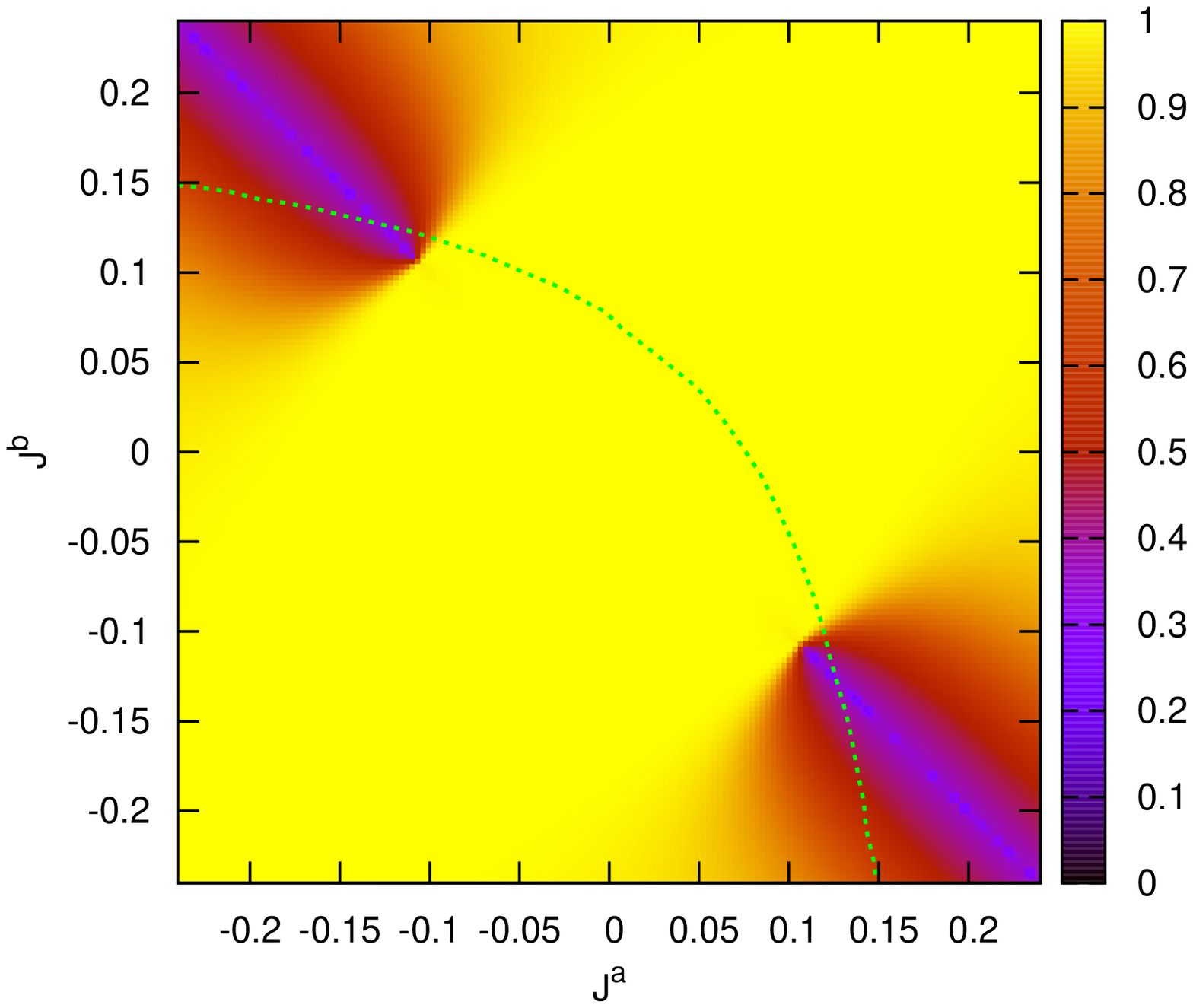}
\caption{\label{Fig1} 
Exact diagonalization results for a square lattice with 2 by 2 sites, at
$I/U=0.45$. The dashed green line in each plot marks the boundary between Mott
and superfluid regime, obtained on the mean-field level as described in Section
\ref{MItrans}.
{\bf Left:} Ground state magnetization with respect to
$S_z=\frac{N_A-N_B}{N_A+N_B}$.
{\bf Middle:} Anticorrelation between nearest neighbors, measured by
$P_{AB}=\sum_{\langle ij \rangle} \langle A_i^\dagger B_{j} + {\rm H.c.}
\rangle$.
{\bf Right:} Overlap of the ground state with a product state, given by Eq.
(\ref{prod}).
}
\end {figure*}

\subsubsection{Limiting cases}
Much of the physics can be understood by considering the limiting cases where $I$ and/or $J_{\pm}
\rightarrow 0$. 

For $J_-=0$ and $I=0$, the Hamiltonian (\ref{spinH}) is equivalent to a
ferromagnetic XXX model, with the states $\ket{AA}$, $\ket{BB}$, and
$\ket{AB}+\ket{BA}$ forming a degenerate ground state manifold. The influence of
a small but finite value of $I$ can be studied by a Taylor expansion: To
first order, $I$ appears only in the term  $\propto I S^z$, that is, it acts as
a magnetic field. In that sense, this term defines the quantization axis of the
ferromagnetic phase, and we accordingly define the magnetization $M$  as the
averaged imbalance between atoms in state $\ket{A}$ and $\ket{B}$:
\begin{align}
\label{mag}
 M = \sum_i \langle S_i^z \rangle / N.
\end{align}
With this definition we see that the states $\ket{AA}$ and $\ket{BB}$ are
oppositely magnetized, $M=\pm1$, while $\ket{AB}+\ket{BA}$ is unmagnetized,
$M=0$. Clearly, at the onset of a small $I$, the degeneracy between these states
is lifted through a linear Zeeman shift, $\Delta E \sim I \Delta M$. The
unique ground state is then the fully magnetized one, $M=1$.

If we next also allow for small but non-zero $J_-$, we have, to lowest order in
$J_-$, to take into account the last term in Eq. (\ref{spinH}), $\propto S^x$. This
term can be considered as an additional magnetic field component, so its
presence will accordingly change the magnetization axis of the system. We thus
expect a ferromagnetic phase with a continuously shifted quantization axis.
With respect to the original quantization axis, this is reflected in a
demagnetization, that is we get ground states with $M<1$.

Another limiting case which can be solved exactly is obtained by
setting $I=0$, while taking $J_-$ and $J_+$ finite. In this case, the ground
state reads $\ket{+} \equiv (\ket{AA}+\ket{BB}+\ket{AB}+\ket{BA})/\sqrt{4}$ if
$J_+ J_- >0$, or $\ket{-} \equiv (\ket{AA}+\ket{BB} -
\ket{AB} - \ket{BA})/\sqrt{4}$ if
$J_+ J_- <0$. Both states are not magnetized, $M=0$. 

An expression which is able to interpolate between all cases discussed
so far is given by the product state ansatz
\begin{align}
\label{prod}
 \ket{\Pi(M)^\pm}_{ij} \equiv \prod_i \left( \sqrt{\frac{1+M}{2}} \ket{A}_i \pm 
\sqrt{\frac{1-M}{2}} \ket{B}_i \right),
\end{align}
where magnetization $M$ is a free parameter. Apparently, independent of $M$,
such ansatz describes ferromagnetic phases. The parameter $M$ accounts for the
shift of the magnetization axis. More generally, one could still introduce a
phase angle between the contributions $\ket{A}_i$ and $\ket{B}_i$. However,
since in Eq. (\ref{spinH}) the magnetic field along $S^y$ is always
zero, the product wave function can be kept real.

A limiting case with very different physical behavior is given by $J_+
\rightarrow 0$. The spin-spin interactions in Eq. (\ref{spinH}) then yield an
XYZ model, with ferromagnetic coupling in the $S^x$-direction, and
antiferromagnetic coupling along the $y-$ and $z$-component of
spin. To study this limit, we first consider the two-site Hamiltonian of Eq.
(\ref{Heff}): The two states $\ket{AB}$, $\ket{BA}$ are decoupled from the
states $\ket{AA}$ and $\ket{BB}$. Their energy is given by
\begin{align}
 E_{\rm AB} = -dN \frac{2J_-^2}{U^2-4I^2},
\end{align}
where the prefactor $dN$ enables to go beyond the two-site picture by counting
the number of nearest-neigbor pairs in a $d$-dimensional system of $N$
particles. The antiferromagnetic states compete with the ferromagnetic states
$\ket{AA}$ and $\ket{BB}$. The ground state energy $E_{\rm FM}$ in the
$\{\ket{AA},\ket{BB}\}$ Hilbert space is found by diagonalizing a two-by-two
matrix:
\begin{align}
-dN\left(
\begin{array}{cc}
\frac{2J_-^2}{U+2I}+\frac{I}{d} & \frac{2J_-^2}{U} \\
\frac{2J_-^2}{U} & \frac{2J_-^2}{U-2I}-\frac{I}{d}
\end{array}
\right)\Psi = E_{\rm FM} \Psi.
\end{align}
By demanding $E_{\rm AB} < E_{\rm FM}$ we obtain the phase boundary between a
ferro- and an antiferromagnetic phase. It is given by:
\begin{align}
\label{afm-boundary}
J_-^{\rm crit}/U = \sqrt{\frac{1}{d}} \sqrt{\frac{1-2(I/U)^2}{2\sqrt{2}}-\frac{1}{4}}.
\end{align}
This formula shows that larger dimensionality extends the antiferromagnetic
regime towards smaller values of $I$ and $J_-$, that is, towards a parameter
regime of best validity of the effective Hamiltonian Eq. ($\ref{Heff}$).

\subsubsection{Exact diagonalization results}

The expectations from analyzing the two limiting cases are supported by a full
numerical solution of the effective Hamiltonian for a small number of particles.
We have considered chains (of up to 12 particles), squares of up to (16
particles), and a cubic arrangement of 8 particles. We calculated the ground
state and low excitations by Lanczos diagonalization of $H_{\rm eff}$, and
evaluated observables like ground state magnetization, defined in Eq.
(\ref{mag}), and spin correlations. The latter we use, in particular, to
identify antiferromagnetic order, which can be tested by an anticorrelation
order parameter 
\begin{eqnarray}
P_{AB} \equiv \sum_{\langle ij \rangle} \langle A_i^\dagger B_{j} + {\rm H.c.}
\rangle.
\end{eqnarray}
This quantity gives the probability of finding a particle in state $B$,
once a neighboring site has been prepared in state $A$, and vice versa. It is
unity only for the two chequerboard states. 

\begin {figure}[t]
\centering
\hspace{0cm}\includegraphics[width=0.23\textwidth, angle=0]{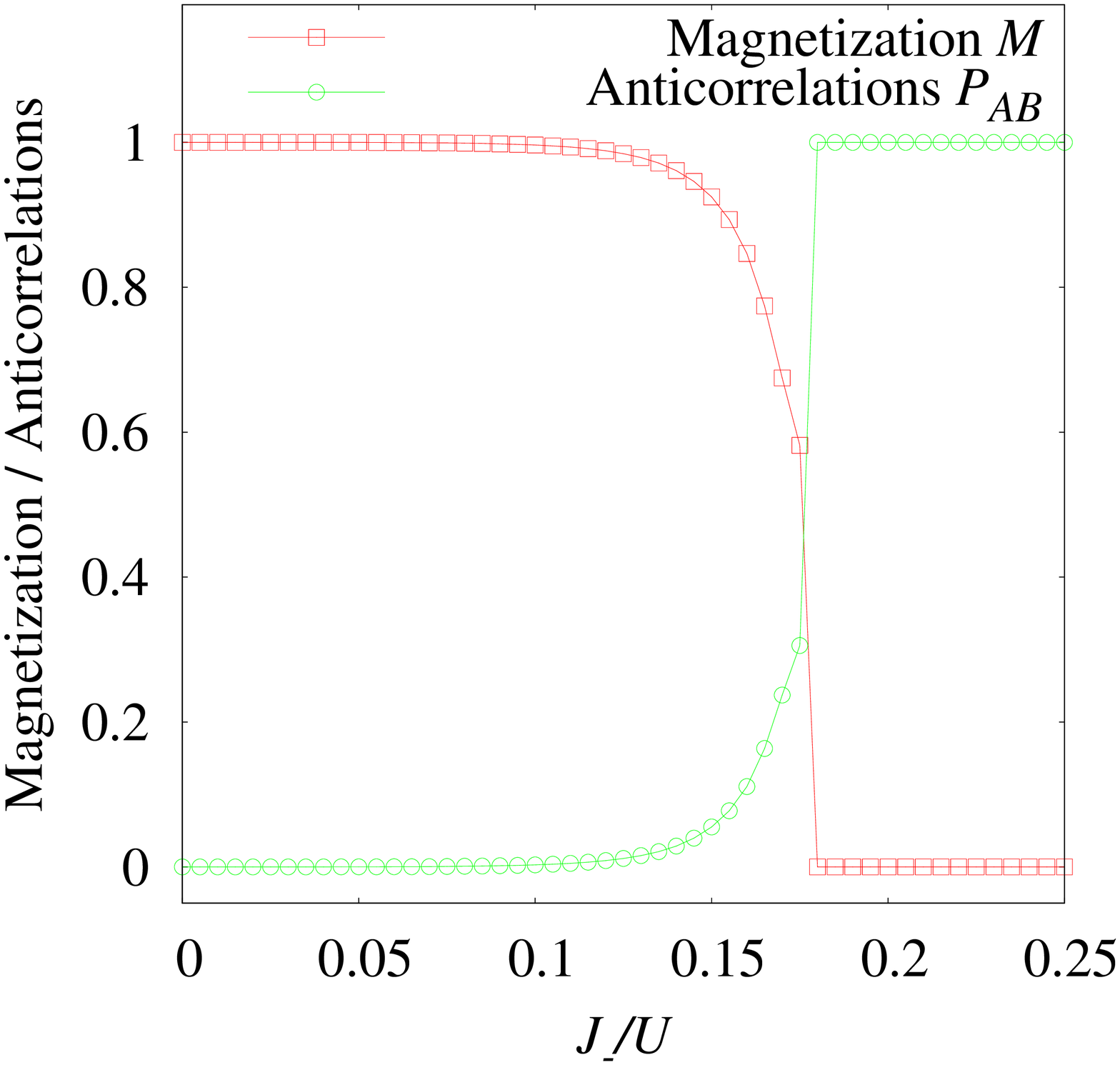}
\hspace{0cm}\includegraphics[width=0.23\textwidth, angle=0]{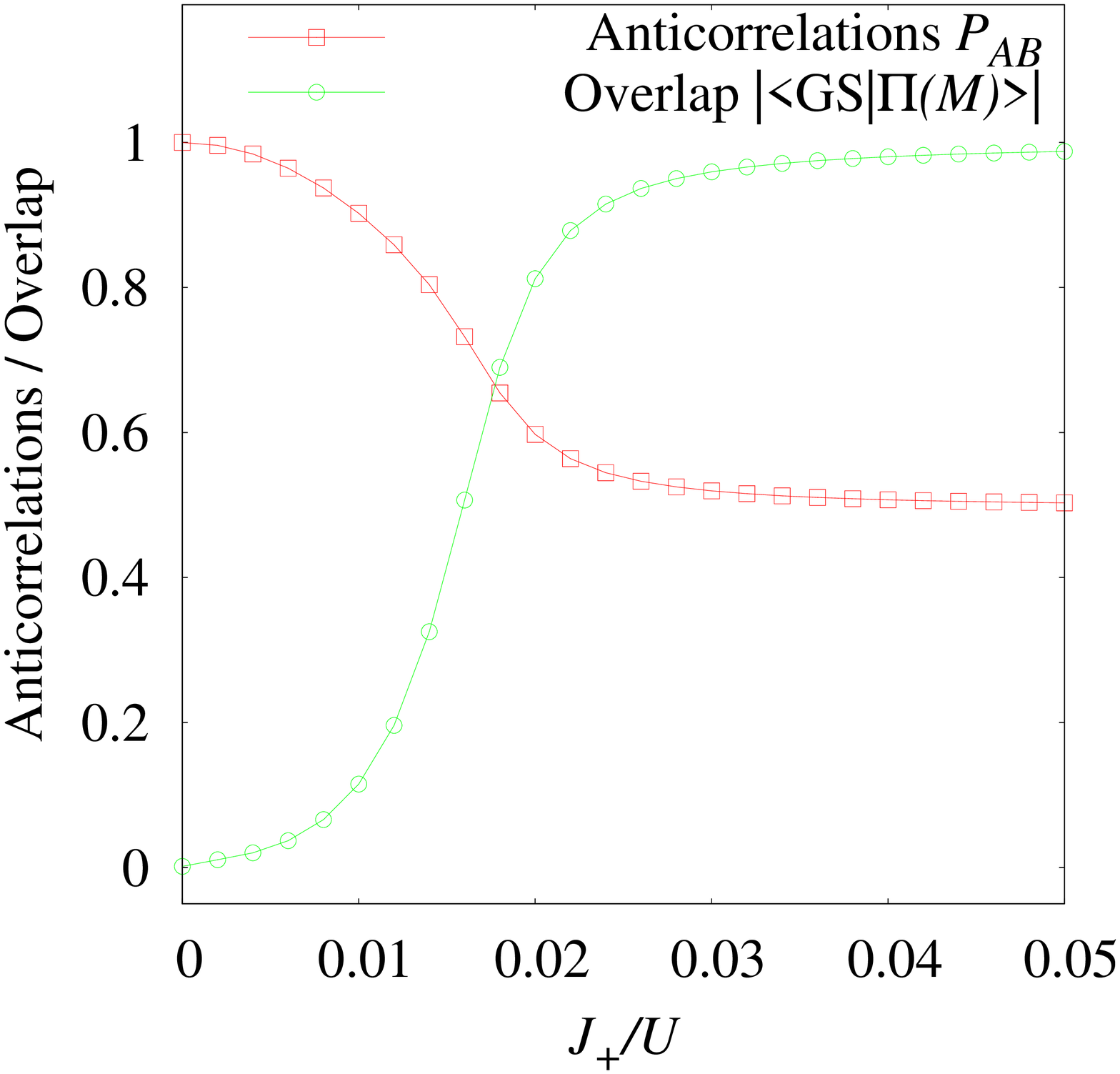}
\caption{\label{Fig2} 
Exact diagonalization results for a square lattice with 4 by 4 sites, at
$I/U=1/3$. 
{\bf Left:} Ground state magnetization $M$ and anticorrelations $P_{AB}$, at
$J_+=0$ as a function of $J_-/U$.
{\bf Right:} Ground state magnetization $M$ and overlap of the ground state with
a product state, given by Eq. (\ref{prod}), at $J_-/U=0.2$ as a function of
$J_+/U$}
\end {figure}

As a test of the ferromagnetic behavior, apart from the magnetization with
respect to $S^z$, we have calculated overlaps between the ground state and
the product state ansatz of Eq.~(\ref{prod}). The results are  shown in Figs.
\ref{Fig1} and \ref{Fig2} for small square lattices. Qualitatively, the same
results are obtained for linear and cubic arrangement. Also the size of the
system does not play a role.  We could not find any qualitative influence of
particle number on the calculated quantities, as long as it is kept
even.

We find a broad regime in which the ferromagnetic solution (with $S^z$
quantization) is the ground state, see  left panel of Fig. \ref{Fig1}. However,
for values $J_-\gtrsim J_+$, the magnetization decreases, but still, as evident
from the right panel of Fig. \ref{Fig1}, the system is in a product state. This
confirms that quantization axis has shifted. We furthermore find that the sign
$\epsilon=\pm$ in Eq. (\ref{prod}) is always defined by the sign of $J_+J_-$,
$\epsilon= \frac{J_+J_-}{|J_+J_-|}$, as expected from the limiting case with
$I=0$. Note that product states with + and - are physically distinguishable in
the original basis of $a$ and $b$ particles. While states
with $M=0$ and $M=\pm 1$ have, at least on average, the same number of $a$ and $b$ particles per site, state with $0<|M|<1$ have an excess of $a$ or $b$
particles, depending on $\epsilon$.

On the line corresponding to $J_+=0$ we find a sharp transition between a state
with relatively large magnetization and small anticorrelations, and the
chequerboard solution, characterized by $M=0$ and $P_{AB}=1$, see left and
middle panel of Fig. \ref{Fig1}, and left panel of Fig. \ref{Fig2}. As a
level crossing is the mechanism behind the transition, the transition is
accompanied by a discontinuity in the first derivative of the energy as a
function of $J_-$. Thus, along the line $J_+=0$, we have a second-order
phase transition.

This is different from the transition which destroys the chequerboard order
through the presence of a non-zero $J_+$. As shown in the right panel of Fig.
\ref{Fig2}, at constant $J_-$ the system smoothly evolves from an
antiferromagnet to a ferromagnet. We note that the regime in which
antiferromagnetic order dominates turns out to be very thin.

It should also be noticed that the chequerboard phase occurs only for
relatively large values of $J_-/U$ and $I/U$. A discussion whether these
parameters still allow for a Mott description will be given in Sec.
\ref{MItrans}. In Fig. \ref{Fig1}, we have anticipated the results from this
section by drawing the Mott boundary (dashed line), obtained from the
mean-field-like calculation presented below. We see that, at least for the
concrete choice of parameters, the antiferromagnetic region coincides partly
with the Mott region.

\subsubsection{Mott-superfluid transition \label{MItrans}}
In our analysis so far we have assumed that the system is in the Mott phase.
For an estimation of the boundary between Mott and superfluid (SF) phase, we
calculate the excitation spectra of the system. The occurrence of a zero-energy
mode signals the breaking of U(1) symmetry, and thus the transition into the SF phase.

The excitation spectra are obtained from the Green functions, which we evaluate
within the first order of a resummed hopping expansion \cite{ednilson,grass-laserphys}.
This approximation is equivalent to a mean-field treatment, which for the standard Bose-Hubbard model is known to give quantitatively good results in $d>1$ dimensions.

Working in imaginary time, and Dirac picture, the time evolution of the
operators reads
\begin{align}
 \hat A(\tau) =& e^{H\tau} \hat A e^{-H\tau}, \\
\hat A^\dagger(\tau) =& e^{H\tau} \hat A^\dagger e^{-H\tau},
\end{align}
with $H=H_I+H_0$. With this we define the ``deep Mott'' (i.e. local) Green
function as
\begin{align} 
\label{green}
G^{(\sigma,\sigma')}(\tau,\tau') =& \sum_{N_A,N_B=0}^{\infty}
\frac{e^{-\beta
E(N_A,N_B)}}{{\cal Z}^{(0)}} \times
\\ \nonumber &
\bra{N_A,N_B} \hat T  \hat O^{(\sigma) \dagger}(\tau) \hat
O^{(\sigma')}(\tau') 
\ket{N_A,N_B},
\end{align}
with $\beta$ the inverse temperature, ${\cal Z}^{(0)}$ the partition function
of the system with $H_J=0$, $\ket{N_A,N_B}$ a Fock state with $N_A$
($N_B$) particles per site in state $A$ ($B$), $\hat T$ the imaginary-time
ordering operator. The index $\sigma$ now stands for $A$ or $B$, and the
operator $\hat O^{(\sigma)}$ is an annihilation operator with respect to the
state $\sigma$. Note that in the $A,B$ basis, the Green function is diagonal.
The object in Eq. (\ref{green}) is most easily evaluated in Matsubara space.
In the limit $\beta \rightarrow \infty$, the Green function reads
\begin{align}
 G_i^{(A,A)}(\omega_{\rm M}) &= \frac{2}{U-\mu-I+i\omega_{\rm M}}-\frac{1}{-\mu-I+i\omega_{\rm M}}, \\
 G_i^{(B,B)}(\omega_{\rm M}) &= \frac{1}{U-\mu-I+i\omega_{\rm M}}.
\end{align}

We can directly apply the formula from first-order resummed hopping
expansion \cite{grass-laserphys,indianpra}:
\begin{align}
 [G_{\bf k}(\omega_{\rm M})^{(\sigma,\sigma')}]^{-1} =& \delta_{\sigma,\sigma'}
[G^{(\sigma,\sigma)}(\omega_{\rm M})]^{-1} - J^{(\sigma,\sigma')} \times
 \\ \nonumber &
[\cos (k_x a) + \cos(k_y a) + \cos(k_z a)] .
\end{align}
Here, we have assumed a cubic lattice, but by neglecting the last cosine, the calculation is also carried out for square lattices.
The poles of this Green function, i.e. the equation $[G_{\bf k}(i \omega)^{(\sigma,\sigma')}]^{-1} =0$, yields the dispersion relations $\omega(\vec{k})$.

In the regime of one particle per site we find three solutions which are gapped and behave quadratically around an extremum at ${\bf k}={\bf 0}$.
Two of these modes are particle modes at positive energy, and the other is a hole excitation at negative energy. 
By increasing the hopping strength, at least one of the solutions becomes gapless, and the system becomes compressible. 
This marks the phase boundary of the Mott phase. For $I=J_-=0$, we obtain the standard (mean-field) Mott lobe. 
At the tip of the lobe, two modes become simultaneously gapless, marking the physical transition point to the superfluid regime. 
The internal hopping $I$ is found to simply shift the first Mott lobe from the
interval $0< \mu/U <1$ to $-I/U < \mu/U <1-I/U$.

\begin {figure}
\centering
\hspace{0cm}\includegraphics[width=0.48\textwidth, angle=0]{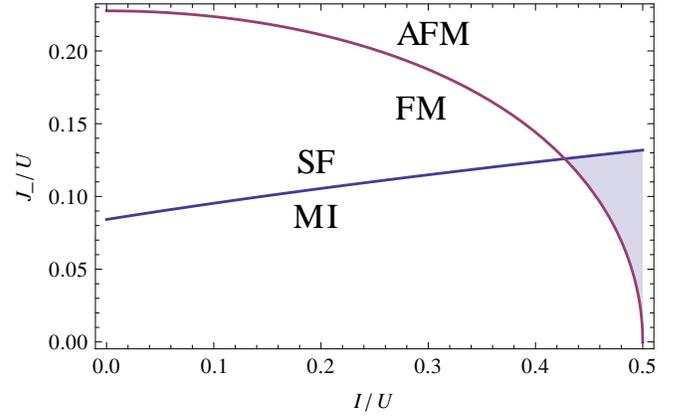}
\caption{\label{Fig3} 
Phase boundaries of a 2D system at $J_+=0$: the blue line marks the (mean-field)
boundary between Mott and superfluid phase. 
The purple line marks the phase boundary between an ferro- and
antiferromagnetically ordered Mott phase. The shaded area marks the region in
which an antiferromagnetic Mott phase is expected.
}
\end {figure}

We are most interested in the regime where $J_+=0$. In this case, a relatively compact expression for the phase boundary is found:
\begin{align}
\label{lobe}
 J_-^{\rm MI/SF}(\mu,I) = \frac{1}{2d} \sqrt{\frac{I-I^3+\mu -2 I \mu -I^2 \mu -2 \mu ^2+I \mu ^2+\mu ^3}{1+I+\mu }}.
\end{align}
Here, all energies are expressed in units of $U$. At fixed $I$, this function
gives the Mott lobe. Interestingly, the Mott boundary scales with dimension as
$1/d$, while the antiferromagnetic boundary behaves as $\sqrt{1/d}$. Therefore,
the parameter region of a possibly antiferromagnetic Mott phase is expected to
be larger in system of less dimensions. Since the mean-field result is not
reliable in 1D, we plot, in Fig. \ref{Fig3}, the tip of the Mott lobe from Eq.
(\ref{lobe}) as a function of $I/U$ for a two-dimensional system. Also the
antiferromagnetic boundary from Eq. (\ref{afm-boundary}) is plotted, and the
shaded region marks a possibly antiferromagnetic Mott regime. It is restricted
to relatively large values of $I/U \gtrsim 0.4$, where the validity of the
effective Hamiltonian Eq. (\ref{Heff}) is doubtable. We have to note, however,
that for the standard Bose-Hubbard model it is well known that the mean-field
calculation underestimates the Mott regime, with an error of around 30\% in two 
dimensions. If this is the case also here, the antiferromagnetic Mott phase
would extend to somewhat smaller values of $I$
, cf. Fig. \ref{Fig3}.

\subsubsection{Experimental detection of the phases}
The bare atomic states $a$ and $b$ can be distinguished by their different
magnetic moment if one applies a real magnetic field along the quantization
axis. In particular, within a time-of-flight expansion through such a field,
this distinguishes between the $\ket{+}$ and the $\ket{-}$ phase. However, this
does not allow to distiguish between the antiferromagnetic and the
ferromagnetic $M=1$ phase, since both possess the same number of
$a$ and $b$ particles. These phases can be distinguished by
applying a magnetic field perpendicular to the quantization axis, with respect
to which the two superpositions $A$ and $B$ posses opposite magnetic moments.

\subsection{Magnetic ordering in SU(${\cal N}$) systems}
We now turn to a discussion of SU(${\cal N}$) systems with ${\cal N}>2$. Due to
the increasing
Hilbert space, a full treatment becomes very tedious. We have seen for
${\cal N}=2$
that the most interesting physics occurs when $j_{\cal N} = J_+ = 0$, such that
$j_1 =
J_-$ is the only contribution to the external hopping. To ${\cal N}>2$, this
generalizes by choosing $j_1=j_{{\cal N}-1}^*$ different from zero, while all
other
hopping processes shall be zero. We further simplify the problem by
choosing $j_1=j_{{\cal N}-1} \equiv j $ to be real. 

With these simplifications, the Hamiltonian can be written as
\begin{align}
\label{Heff3}
 H = - j \sum_{\langle ij \rangle,\sigma} \left( \hat A_j^{({\rm mod }
[\sigma\pm 1,{\cal N}])\dagger} \hat A_i^{(\sigma)} +{\rm H.c.} \right) + H_0.
\end{align}
Here, every spatial hopping is connected to a hopping in the internal dimension.
In SU(3), such a hopping term can be achieved by making $J^{(2)}=J^{(3)}$ and
$J^{(1)}= -2J^{(2)}$. In SU(4), we have to make $J^{(1)}=-J^{(3)}$ and
$J^{(2)}=J^{(4)}=0$.

Assuming again the Mott limit of one atom per site, we consider the
effective Hamiltonians for the double-well problem. Let us denote the
eigenstates of $H_I$ as $A,B$ for SU(2), $A,B,C$ for SU(3), and $A,B,C,D$ for
SU(4). As explained before, the SU(2) system has a degeneracy between the state
$\ket{AB}$ (that is with the left atom in $A$ and the right atom in $B$),
$\ket{BA}$, and $(\ket{AA}+\ket{BB})/\sqrt{2}$. This degeneracy is a
consequence of the fact that $\ket{AB}$ and $\ket{BA}$ (or more generally: the
two checkerboard states) are already eigenstates of a hopping which changes the
pseudospin state by one. In contrast to this, a checkerboard solution
does not exist for the SU(3) system. Accordingly, we find a unique ground state
given by
$(\ket{AA}+\ket{BB}+\ket{CC}+\ket{AC}+\ket{CA}+\ket{AB}+\ket{BA}+\ket{BC}+\ket{
CB})/3$.

The physics of the SU(4) system reduces to the SU(2) physics with three
degenerate ground states in the two-site limit. This is not very
surprising if we note that the Hamiltonian (\ref{Heff3}) is obtained from the
original Hamiltonian $H_I+H_J+H_0$ by setting the hopping strength of two
components to zero. In order to make the connection to the SU(2) case clearest,
we define the local states $\ket{a}=(\ket{A}+\ket{C})/\sqrt{2}$ and
$\ket{b}=(\ket{B}+\ket{D})/\sqrt{2}$. In terms of these states, the three
degenerate ground states of the effective SU(4) Hamiltonian read $\ket{ab}$,
$\ket{ba}$, and $(\ket{aa}+\ket{bb})/\sqrt{2}$.

As already noted in Sec.~\ref{system}, for the right choice of $\phi$ in Eq.
(\ref{HI}) it is possible
to make the internal energies of the laser coupling, given by Eq.
(\ref{lambda}), equidistant. As argued before, this energy, divided by the
coupling strength $I$, is somewhat similar to a magnetization of the state. We
then find that $\ket{a}$ and $\ket{b}$ are states with opposite magnetization,
just as the states $\ket{A}$ and $\ket{B}$ in the SU(2) case. Accordingly, the
degeneracy will be lifted by $I$ in the same manner as discussed before, namely
through a quadratic Zeeman shift.

\section{SU(${\cal N}$) symmetry breaking in the interaction}\label{model}

Let us now discuss the case of $u\neq 1$, i.e. of the breaking of the SU$({\cal N})$-invariance already 
at the level of interaction. In particular, we focus on the region of parameters
where 
$H_J$ couples the degenerate minima 
 dominantly at second order in perturbation theory. 

As derived and discussed in full detail in the Appendix \ref{mott},  
for $0<U/U_d<1$ and for any $q\in {\mathbb N}$ it exists a
non-zero window for the chemical potential, $\mu_1<\mu<\mu_2$, where the
low-energy Hilbert space of the Mott phase is spanned by the states
with $q+1$ particles in one species and $q$ in all the others. In formulas,
these ${\cal N}$ normalized states are  $\ket{\bar\sigma}\equiv
\frac{{a^{\bar\sigma}}^\dagger }{\sqrt{q+1}}\ket {q}$, where $\ket {q}\equiv
\sum_\sigma \frac{({a^{\sigma}}^\dagger )^q}{\sqrt{q\!}}\ket {0}$ is the state
of $q$ bosons for each species. 
This case is interesting as the perturbation theory in the spatial and species
hopping induces a novel Potts-like effective Hamiltonian that displays different
quantum phases.

\subsection{Effective Hamiltonian}

Indeed, it is immediate to see that the hopping term in $I$ connects already
at first order two vectors of the minima's subspace  while the term in $J$ can
only act at second order in perturbation theory. This implies that interesting
physics may appear within the range of validity of the effective Hamiltonian
when $I/U$ and $J^2/U^2$ are of the same order of magnitude and much smaller
than 1. 
As the Hilbert space where the effective Hamiltonian acts is  ${\cal N}^{N_L}$
degenerate, $N_L\equiv$ number of sites, and it is generated by product states
$\{\ket{\bar{\bf\sigma}}\equiv \prod_i\ket {\bar\sigma(i)} \}$, $H_{eff}$ can be
written in terms of the matrix elements over such states
\begin{equation}
H_{eff} = \sum_{\bar{\bf \sigma}\bar{\bf \rho}} \ket {\bar{\bf \rho}} \bra
{\bar{\bf \rho}}H\ket {\bar{\bf \sigma}} \bra {\bar{\bf \sigma}}= \sum_{\bar{\bf
\sigma}\bar{\bf \rho}} \ket {\bar{\bf \rho}} H_{\bar\rho\bar\sigma} \bra
{\bar{\bf \sigma}}\,.
\end{equation}
By denoting the creation (annihilation) operators in the neighborhood of the site $i$ as $a^\dagger_{il}$ ($a_{il}$), $l=1,\dots,2d$, the familiar expression for the matrix element is
\begin{multline}
H_{\bar\rho\bar\sigma}= -
\sum_i\left(J^2\sum_{\bar{\bf\gamma},l,\sigma^\prime,\rho^\prime}\frac{\bra
{\bar{\bf \rho}} a_i^{\sigma^\prime \dagger} a_{il}^{\sigma^\prime}\ket
{\bar{\bf \gamma}}\bra {\bar{\bf \gamma}} a_i^{\rho^\prime \dagger}
a_{il}^{\rho^\prime}\ket {\bar{\bf \sigma}}}{ \Delta_\gamma}\right.\\ \left.
+I  \sum_{\sigma^\prime}\bra {\bar{\bf \rho}}\left(a_i^{\sigma^\prime \dagger}
a_{i}^{\sigma^\prime+1}+a_i^{\sigma^\prime \dagger}
a_{i}^{\sigma^\prime-1}\right)\ket {\bar{\bf \sigma}}\right),
\end{multline}
where $\ket{\bar{\bf \gamma}}$ is any excited state obtained from a minimal energy
state by acting with the hopping term on one link, and $\Delta_\gamma$ is the
energy gap of such state. It is worth to notice that there are only four kinds
of excitations. Indeed, there are two ways of adding a particle on one site:
having two species with $q+1$ particles and the rest with $q$, let us call it
$(++)$ configuration, or having one species with $q+2$ particles and rest with
$q$, a $(2+)$ configuration. On parallel, when one particle is removed, we have
a $q$ configuration, $q$ particles in each species, or a $(+-)$ configuration,
with occupation $q+1,q-1,q\dots q$. In this language, the minimal energy
configuration with $q+1$ particle in one species and $q$ for the rest is a $(+)$
configuration. Altogether, the four possibility are $\gamma_{++-+}$,
$\gamma_{++q}$, $\gamma_{2+-+}$ and $\gamma_{2+q}$ (with the constraint that the
$-$ occurs for  the same species as the $+$ in the other site of the link), with
the 
corresponding gap measured in units of $U$:
\begin{align}
&\Delta_{++-+}= 1,\ \Delta_{++q}= u,\\ 
&\Delta_{2+-+}= 2 - u,\ \Delta_{2+q}=1\,.
\end{align}
 
After some simple  but lengthy calculation, one finds   
\begin{align}
H_{\bar\rho\bar\sigma}
=  & -J^2 \left(2 N_L  d\,  A + B \sum_i\sum_{l=1}^{2d}
\delta^{\bar\sigma(i)\bar\sigma(il)}\right) \mathds{1} -
\nonumber \\ &
 \sum_i\Bigg( J^2 C
\sum_{l=1}^{2d} 
\delta^{\bar\sigma(i)\bar\rho(il)}\delta^{\bar\rho(i)\bar\sigma(il)} +
I(q+1)\times
\nonumber \\ &
\left(\delta^{\bar\sigma(i)(\bar\rho(i)+1)}+\delta^{
\bar\sigma(i)(\bar\rho(i)-1)} \right)\Bigg)\prod_{i'\neq i,il}
\delta^{\bar\sigma(i')\bar\rho(i')}\,. \label{heff}
\end{align}     
 The coefficients $A$, $B$ and $C$ are functions of the coupling $u\equiv \frac
{U_d}U$, the number of species ${\cal N}$ and the total number of particles in
each site $ q{\cal N} +1$
\begin{align}
A&= \frac{q(q+1)({\cal N}-2)}{\Delta_{++-+}}+ \frac{(q+1)^2}{\Delta_{++q}} +
\frac{q(q+2)}{\Delta_{2+-+}}\cr
 &= q(q+1)({\cal N}-2)+ \frac{q(q+2)}{2-u}+ \frac{(q+1)^2}u,\\
B&= \frac{q(q+1)}{\Delta_{++-+}}-2 \frac{(q+1)^2}{\Delta_{++q}} -
\frac{q(q+2)}{\Delta_{2+-+}}+ \frac{(q+1)(q+2)}{\Delta_{2+q}} \cr
 &=-\left(2 (q+1)^2 \frac{1 - u}u + \frac {q (q+2)}{2 - u}\right),\\
C&= \frac{(q+1)^2}{\Delta_{++q}}= \frac{(q+1)^2}u.     
\end{align}   

Few comments on the above coefficients are in order. $A$ is just a constant
shift of the energy and is always positive for the relevant region of the
parameters' space, ${\cal N}\ge 2$, $q\ge0$ and $0<u<1$. The non-trivial part of
the Hamiltonian is determined by the coefficient $B$ and $C$. $B$ is always
negative in the relevant region of parameters' space, and, with the exclusion of
a small corner around $u\sim 1$, its modulus is bigger than $C$, which is always
positive. In particular, in all parameters' region it holds $|B|/C<2$: this has
dramatic consequences on the nature of the ground-state, as it will be discussed
in Sec. \ref{gutz}.  

 As a natural consequence of the periodic identification of the species, modulo
${\cal N}$, the effective Hamiltonian is $\mathbb Z^{\cal N}$ invariant. The
implications of such invariance can be made transparent by rewriting the
effective Hamiltonian in terms of operators. Following the notation of $\mathbb
Z^{\cal N}$ lattice gauge theory \cite{Tagliacozzo2012}, we introduce the unitary operators $X$ and $Z$
 \begin{align}
 &X\ket{\sigma}=e^{i \frac{2 \pi}{\cal N} \sigma}\ket{\sigma} \rightarrow
XX^\dagger=\mathds{1}  \\
 &ZX=e^{i\frac{2\pi}{\cal N}}XZ \rightarrow Z\ket{\sigma}=\ket{\sigma-1}, \
Z^\dagger\ket{\sigma}=\ket{\sigma+1}\,.
 \end{align}
 
 By representing the states $\ket{\sigma}$ as unit vectors ${\bf v}_\sigma$ of
components $v_{\sigma}^a={\delta_{\sigma}}^a$, the operators $X$ and $Z$
correspond to the ${\cal N}\times{\cal N}$ matrices
 $X=\rm{diag}\{e^{i\frac{2\pi}{\cal N}},e^{i\frac{4\pi}{\cal
N}},\dots,e^{i\frac{2({\cal N}-1)\pi}{\cal N}},1\}$ and $Z_{ab}=\delta_{a{\cal
N}}\delta_{b1}+ \delta_{(a+1)b}$.
 
 In order to implement the Hamiltonian Eq. \ref{heff}, in terms of $X$ and $Z$,
there are two main obstacles. The first is to construct the projector $P_{il}
\ket{\sigma_i}\otimes\ket{\sigma_{il}}=\delta_{\sigma_i\sigma_{il}}\ket{\sigma_i
}\otimes\ket{\sigma_{il}}$. Such an operator can be a function of $X_i$ and
$X_{il}$ only, as it does not change the species. By noticing that $X_i\otimes
X_{il}^\dagger \ket{\sigma_i}\otimes\ket{\sigma_{il}}=e^{i\frac{2\pi}{\cal N}
(\sigma_i-\sigma_{il})} \ket{\sigma_i}\otimes\ket{\sigma_{il}}$, it is easy to
verify that $P_{il}=\frac 1{\cal N} \sum_{m=1}^{\cal N} (X_i\otimes X_{il})^m$.
The second is the implementation of the exchange operator $R_{il}
\ket{\sigma_i}\otimes\ket{\sigma_{il}}=  
\ket{\sigma_{il}}\otimes\ket{\sigma_i}$. Its action can be obtained by rotating
simultaneously the spins  using  $Z_{i}\otimes Z^\dagger_{il}$ to the power
$m=\sigma_i-\sigma_{il}$. Formally, it can be achieved for any
$\sigma_i$ and $\sigma_{il}$ using the projector $P_{il}$, $R_{
il}=  \sum_{n=1}^{\cal N}  (Z_{i})^n\otimes\mathds{1}_{il}\cdot P_{il}\cdot
\mathds{1}_i\otimes (Z^\dagger_{il})^n$. As the term in $I$ is trivial, the
Hamiltonian in the operator fashion is
 \begin{align}
 H_{eff}= - \sum_i\left(J^2\sum_{l=1}^{2d} \left(B P_{il}+ C
R_{il}\right) + I \left(Z_i+Z_i^\dagger\right)\right).\label{heffop}
 \end{align}

 \subsection{Exploring the phase space with a  Gutzwiller ansatz}\label{gutz}
 
The ground state of the effective Hamiltonian (\ref{heff}) can be computed in
the mean-field approximation using the Gutzwiller ansatz $\ket\Psi=\prod_i
\ket{\Psi_i}$ with $\ket{\Psi_i}=\sum_{\sigma} f_i^\sigma\frac{a^{\dagger
\sigma}}{\sqrt{q+1}} \ket q$, $\sum_\sigma |f_i^\sigma|^2=1$. As usual, the
$f_i^\sigma$ are variational parameters that are determined by minimizing the
expectation value of the effective Hamiltonian on $\ket\Psi$
\begin{align}
& \bra\Psi H_{eff}\ket\Psi = \sum_{\{\bar\sigma\}\{\bar\rho\}}
\bra\Psi\ket{\bar\rho}\bra{\bar\rho}H_{\bar\rho\bar\sigma}\ket{\bar\sigma}\bra{
\bar\sigma}\ket\Psi=
\nonumber \\ &
={\rm const.} - \sum_{i,\sigma}\Bigg(J^2 \sum_{l,\rho}
\left[f_i^{*\rho}
f_{il}^{*\sigma}\left(B f_i^{\sigma} f_{il}^{\rho}\delta_{\sigma\rho}
+Cf_i^\sigma f_{il}^\rho\right)\right] +
\nonumber \\ & 
(q+1)I \,\left[ e^{i\phi} f_i^{*\sigma}f_i^{\sigma+1} + e^{-i\phi}
f_i^{*\sigma}f_i^{\sigma-1} \right] +\lambda^i\left(\left|f_i^\sigma\right|^2 -
1\right)\Bigg),\label{hgutz}
\end{align}
where with an abuse of notation we include the Lagrange multipliers $\lambda^i$
to variationally impose the normalization of $\ket\Psi$ to one.

However, the main features of the ground-state of the effective Hamiltonian in
the
mean-field approximation can be understood without calculations. In particular,
the phase diagram of the model can be read off by comparing the different terms
that are competing and the relative strength of the corresponding coupling
constant $B$, $C$ (which are controlled by $u$ and ${\cal N}$), and $I/J^2$. 
Let us start by describing the term sourced by $B$. Within the Gutzwiller ansatz
it can written as $H_B\equiv\sum_{il\sigma}|f_i^\sigma|^2|f_{il}^\sigma|^2$,
where the negative sign in the expectation value Eq.~(\ref{hgutz}) is cancelled
by
the negative sign of $B$. As $H_B$ is positive definite its minimum corresponds
to zero. It is evident that a checker-board-like configuration annihilates
$H_B$, for instance $f_i^\sigma=\delta^{1\sigma}$,
and $f_{il}^\sigma=\delta^{2\sigma}$. Such configuration can be consistently
extended to the whole hypercubic lattice as it is bipartite. This tells
us that whatever such term in
the Hamiltonian dominates over the others the ground-state is not translational
invariant.
Now, let us consider the term sourced by $C$. This time we have $H_C\equiv
-\sum_{il} |\vec f_i^* \vec f_{il}|^2$. As the $\vec f$ are normalized to 1, the
modulus square of the scalar product can be at maximum 1, when the vectors are
parallel or anti-parallel. This implies that the minimization of $H_C$ appoints
to translational invariant configurations. 
In view of the above considerations, it is crucial to determine which of the two
terms, and under which conditions, wins over the other. It is worth to notice
that if translational invariance of the ground-state is assumed, this
corresponds to uniform superposition  of $\ket{\sigma}$ states, i.e
$|f^\sigma|=1/\sqrt{\cal N}$ for any  $\sigma=1,\dots,{\cal N}$,
as obtained analytically by minimizing the fourth order polynomial. As the
phases
of the $f^\sigma$ are irrelevant we may choose for simplicity
$f^\sigma=1/\sqrt{\cal N}$.
Such observation allows us to derive an not rigorous argument to decide which is
the actual ground-state depending on the relative values of $B$ and $C$ when
$I=0$.        
The total energy per site of the translational invariant configuration above,
$\ket\Phi=\frac 1{\sqrt{\cal N}} \ket{\sigma}$, will be $E_\Phi=B/{\cal N}- C$.
As a checker-board configuration has zero total energy, this means that $\ket
\Phi$ is favorable whatever $E_\Phi<0$. For our effective Hamiltonian Eq.
\ref{heff} this condition is always realized as $-B/C<2$. Hence, we conclude
that the ground-state is translational invariant for $I=0$, at any value of
${\cal N}\ge 2$, $q\ge 0$ and $0<u<1$.
The above argument is it not rigorous as a priori we cannot exclude the
existence of less energetic and non-translational invariant configuration which
it is not a minimum of $H_B$ and $H_C$ for separated, but the numerical evidence
seems to discard such possibility.  

Now, let turn our attention to the term source by $I$. As this is a local
term, it is straightforward to find the translational invariant ground-state it
selects. Depending on the value of $\phi$, it is a certain eigenvector of the circulant
matrix,
$(ZZ^\dagger)_{\sigma\sigma^\prime}=\delta_{\sigma(\sigma^\prime-1)}+\delta_{
\sigma(\sigma^\prime+1)}$, where ${\cal N}+1$ is identified with 1. For instance, for $\phi=0$ and $\phi=\pi$, it is the one 
corresponding to its maximal and minimal eigenvalues, respectively. 
As maximal eigenvalue and the corresponding
eigenvector  for any number of species are 2 and $f^\sigma=1/\sqrt{\cal N}$ for
any {\footnotesize $\sigma=1,\dots,{\cal N}$}, respectively, the ground-state at $\phi=0$ is
$\ket{\Phi}$ for any $I$ (by definition $I>0$). 

  In fact, the situation is more interesting for $\phi=\pi$, or for any other value providing a degeneracy of two eigenvalues. The former is equivalent to reverse the sign of $I$, and the
eigenvector corresponding to the minimal eigenvalue  has a non-trivial
dependence on ${\cal N}$. In particular, for ${\cal N}$ odd there are two
degenerate states with minimal intra-species hopping term,
$\lambda_{min}=-2\cos(\pi/{\cal N})$, while for ${\cal N}$ even the state is
unique, $\ket \Phi_{alt}=\frac 1{\sqrt{\cal N}} \sum_{\sigma}
(-1)^\sigma\frac{a^{\dagger \sigma}}{\sqrt{q+1}} \ket q$, with
$\lambda_{min}=-1$. 
It is worth to notice that in the $d+1$ picture, where the different species
correspond to layers in the {\it compact} extra-dimension (as the hopping in
$I$ acts as a circular matrix where ${\cal N}+1$ is identified with 1), a
negative $I$ implies the presence of a magnetic $\pi$-flux in the compact
dimension.

In view of the above consideration, the existence of at least one phase
transition is predicted within the mean-field approximation 
 due to the hopping term between species. Indeed, starting with  $I$ positive
and decreasing it to negative values, the stable translational invariant phase
determined by  
$\ket\Phi$, becomes metastable and a new translational invariant state takes
over. If ${\cal N}$ is even, the ground-state is simply determined by $\ket
\Phi_{alt}$. For ${\cal N}$ odd the situation is slightly more involved:
The
minimal local state is given by the linear combination of the two minimal
eigenvectors of the circulant matrix that minimizes the effective Hamiltonian at
$I=0$.

At this stage we cannot exclude the existence of other intermediate phases and
corresponding phase transition. However, the above scenario is confirmed by
numerical evidence.
Let us analyze the results for even and odd numbers of species separately.
In the former case, ${\cal N}=2r$, the state $\ket{\Phi}$ and $\ket{\Phi_{alt}}$
have the same energy contribution from $H_B$ and $H_C$, the transition happens
exactly at $I=0$, for any value of $d$, $r$, $q$ and $u$.  

In the latter, ${\cal N}=2r+1$, the main difference is that $\ket{\Phi_{alt}}$ is replaced 
by the linear combination of two minimal eigenvectors of $ZZ^\dagger$ which minimize $H_B$ and $H_C$.
In fact, contrary to the SU$({\cal N})$-symmetric case, $u=1$, the interaction part of the Hamiltonian $H_0$
is not invariant under a change of basis to the eigenvalues of $H_I$ for generic $u$.

We conclude with two key observations.
First, as in the SU$({\cal N})$-invariant case, checkerboard-like solutions can be achieved by considering species-dependent hopping term $H_J$.
Second, again for any values of $u$, the different states and phases can be distinguished by time of flight experiment combined with a Stern-Gerlach one.

\section{Conclusion and Outlook}

One of the major results we present here is that 
SU$({\cal N})$-breaking spatial hopping together with 
species mixing terms can induce inhomogeneous phases, i.e. the magnetization displays 
a crystal structure. Note that 
in the extradimension picture, in which the different pseudo-spin states become
different sites, such pattern becomes a density pattern.
Hence, the existence of Mott states of such sort suggests the presence of
superfluid phases respecting the same crystal structure, i.e.
supersolid phases. Such appearance is not totally surprising as the SU$({\cal N})$-symmetric interaction  is long-range in the extradimension picture. 
In fact, a practical advantage of the synthetic dimension is that the long-range interaction is naturally not small.
It is worth to notice that such scenarios naturally extend to approximately
SU$({\cal N})$-symmetric interactions, which is more realistic for bosons.

\acknowledgments{
We acknowledge enlightening discussions with J. Rodriguez Laguna and L.
Tagliacozzo and support from  ERC Advanced Grants
QUAGATUA and OSYRIS, EU IP SIQS, EU STREP EQuaM, Spanish MINCIN (FIS2008-00784
TOQATA), and Fundaci{\'o} Cellex. 
}

\appendix

\section{Minimal energy ``Mott" states}\label{mott}

For $J=I=0$, the Hamiltonian  reduces to $H_0$, i.e.to the sum of local
terms $H_0= U \sum_i h_{mott}(i)$ with
\begin{equation}
h_{mott}(i)=\sum_{\sigma} \left(-m \hat n_i^{\sigma} +\frac 12
(\hat n_i^{\sigma})^2\right)+ \frac{u}2 \sum_{i,\sigma\sigma^\prime}
\hat n_i^{\sigma}\hat n_i^{\sigma^\prime},\label{hmott}
\end{equation}
where $m\equiv\mu/U+1/2$ and $u\equiv U_d/U$. As minimizing $H_{mott}$ is
equivalent to minimize each $h_{mott}(i)$, in what follows we omit for brevity
the position index $i$. As $[h_{mott},\hat n^\sigma]=0$, we can minimize on the
occupation basis and the number operators as numbers. Hence, it is convenient to
identify the occupations of each species as  the components of ${\cal N}$-vector
${\bf v}$, $v^\sigma=\hat n^\sigma$ and to express the quantity to be minimize as the
algebraic problem:
\begin{equation}
h_{mott}= \frac 12 {\bf v}^{\rm t}\cdot M \cdot {\bf v} -{\bf w}\cdot {\bf
v}\,, 
\end{equation}
where $M_{\sigma\rho}=u+(1-u)\delta_{\sigma\rho}$, and $w_\sigma= m$, 
{\footnotesize $\sigma,\rho=1,\dots{\cal N}$}.

For any ${\cal N}$, $M$ is a real symmetric matrix that can be diagonalized to
$diag\{1+({\cal N}-1)u,1-u,\dots,1-u\}$ in the following orthonormal basis 
\begin{align}
& {\bf e}_1 = \frac 1{\sqrt{\cal N}}(1,\dots,1) \nonumber \\
&{\bf e}_s=(\overbrace{-\frac 1{\sqrt
{s(s-1)}},\dots}^{s-1},\sqrt{\frac{s-1}s},\overbrace{0,\dots}^{{\cal N}-s}),\
\text{\footnotesize $s=2,\dots,{\cal N}$}.
\end{align}

This implies that $h_{mott}$ becomes the sum of quadratic functions of each of
the variables 
\begin{align}
& X^1\equiv {\bf v}\cdot {\bf e}_1= \frac{\sum_{\sigma=1}^{\cal N}
\hat n^\sigma}{\sqrt{\cal N}},\nonumber \\
& X^s\equiv {\bf v}\cdot {\bf e}_s=-\frac{\sum_{\sigma=1}^{s-1}
\hat n^\sigma}{\sqrt{s(s-1)}} + \sqrt{\frac{s-1}s} \hat n^s. 
\end{align}
By introducing $\nu=\sum_{\sigma=1}^{\cal N} \hat n^\sigma$ to indicate the total
number of particle per site, we have
\begin{equation}
h_{mott}= \frac{1+({\cal N}-1)u}{2{\cal N}} \nu^2- m\nu +
\frac{1-u}2\sum_{s=2}^{\cal N } (X^s)^2\,.
\end{equation}
It is straightforward to minimized the above expression, at least when $\nu$ is
commensurable with number of species. Depending on the sign of the quadratic
terms, we can distinguish 3 main regions. For $u<-1/({\cal N}-1)$, $h_{mott}$ is
unbounded from the below as the system tries to acquire as many particles as
possible for any value of the chemical potential. Precisely at $u=-1/({\cal
N}-1)$, the energy of the system is not depending on the number of particles per
site but only on their distribution between the different species. 

For $-1/({\cal N}-1)<u<1$, both the terms in $\nu$ and $X$s are convex. If $m$
is such that $\nu= q {\cal N}$, for $q\in {\mathbb N}$, $m=m_q\equiv (1+({\cal
N}-1)u) q$, the minimum is achieved at $X^s=0$ for any {\footnotesize
$s=2,\dots,{\cal N}$}, i.e. for a uniform distribution of particles
$\hat n^\sigma=q$, {\footnotesize $\sigma=1,\dots,{\cal N}$}. For a generic value of
$m$, the terms in $\nu$ and $X$s compete. It can be shown that for $-1/({\cal
N}-1)<u<0$ the system admits only the commensurable filling $\nu= q {\cal N}$
with uniform distribution per species, while for $0<u<1$ the system explores
configurations with exceeding particles over the uniform distribution. In
practice, if we increase $m$ starting from $m_q$, in the latter case the
configuration with one particle added in any one of the species   becomes less
energetic than uniform configuration before $m$ reaches $m_{q+1}$ while in the
former case it does not occur. Indeed, by relabeling the species to have the
extra particle in last one, 
$X^s=0$, {\footnotesize $s=2,\dots,{\cal N}-1$}, $X^{\cal N}=\sqrt{({\cal
N}-1)/{\cal N}}$, and  the energy of such configuration is 
\begin{align}
E_{q,1}&=\frac{1+({\cal N}-1)u}{2{\cal N}} (q{\cal N} +1)^2- m(q{\cal
N}+1)\nonumber \\
       &\ \ \ \ + \frac{(1-u)({\cal N}-1)}{2{\cal N}}\nonumber \\  
   &=E_q+\Delta_{q,1},\\
\text{to be }&\text{compared with,} \nonumber \\
E_{q+1}&= (\frac{1+({\cal N}-1)u}2 (q+1)- m) (q+1) {\cal N} \nonumber \\
       &=E_q+\Delta_{q,{\cal N}},
\end{align}
where $E_q=(\frac{1+({\cal N}-1)u}2 q- m) q {\cal N}$ is the energy of the
uniform distribution with $q$ particle per species, and $\Delta_{q,1}= -m +m_q
+1/2$ and $\Delta_{q,{\cal N}}=(1/2 (1+({\cal N} -1))(2 q +1)-m) {\cal N}$  are
the gaps with respect to it of the configuration with an extra particle, and of
the uniform distribution with $q+1$ particles per species, respectively. Hence,
by imposing $\Delta_{q,1}\le0$ while  $\Delta_{q,{\cal N}}>0$ for ${\cal N}>2$,
the condition $0<u<1$ is found. 
The above statement can be rigorously proved by consider all the minimal energy
configurations with total on-site occupation $q{\cal N} \le\nu\le (q+1){\cal
N}$. Indeed, it can be shown that the energy associated to them is
\begin{equation}
E_{q,p}=\frac{1+({\cal N}-1)u}{2{\cal N}} (q{\cal N}+p)^2 - m(q{\cal N}+p)+
\frac {1-u}2 p\frac{{\cal N}-p}{\cal N},
\end{equation}   
where $E_{q,0}=E_q$ and $E_{q,p}=E_{q+1}$. Accordingly, the gap
\begin{align}
\Delta_p&=E_{q,p}-E_q \nonumber \\
        &=p\left(\frac{1+({\cal N}-1)u}{2{\cal N}} (2q{\cal N}+p) + \frac {1-u}2
\frac{{\cal N}-p}{\cal N} - m\right),
\end{align}   
turns out to be  monotonically growing function of $p$ for $0<u<1$ and to be
monotonically decreasing function of $p$ for $-1/({\cal N}-1)<u<0$.  

The point $u=1$ separating the second region from the third region is special as
it represents the SU$({\cal N})$ symmetric point. The energy simply depends on
the $\nu$ and the ground state is as degenerated as all the possible ways of
distributing $\nu$ particles in ${\cal N}$ boxes.

The third region, $u>1$, the quadratic term in the total occupation is positive
definite while the one in the $X$s is negative definite, hence the minimum of
the energy is obtained by maximizing the $X$s for fixed $\nu$. It is immediate
to realize  that this happens then all the particles seats in one species. 

\subsection{The ground-state degeneration and the effective Hamiltonian in
perturbation theory}
     
As our strategy is to find novel many-body effect with the reach of an effective
theory approach, the most interesting regions of parameters are the ones
displaying a degenerate ground-state. In particular, to be the effective
Hamiltonian of physical significance, the hopping terms in $J$ and $I$ should
act not trivially on the minimal energy subspace at low order in perturbation
theory, let us say at most at second order. From the analysis of the previous
section, the degeneration of the minima it is present for $u=-1/(1+({\cal N}-1)$
and $u>0$. In the former case, the situation in presence of just two species,
${\cal N}=2$, has been already studied in \cite{Trefzger09}. For a generic
${\cal N}$, the spatial hopping terms can act differently than the identity only
at order ${\cal N}$, as this is the minimal number of particles to be moved and
each hopping operation can move one. 

In the latter case, similar reasoning applies when $u>1$ but this time the
quantity that determines the order of perturbation is the on-site occupation.
The most interesting case is for $0<u\le 1$. Let us focus before on $0<u<1$.
Here, as shown in previous section,  for any positive integer $q$ and any $p$
between $1$ and ${\cal N}-1$ it exists a value of $m$ such that the minimal
energy configurations have $\nu= q{\cal N} +p$ and correspond to ${\cal N}-p$
species populated by $q$ particles and $p$ species populated by $q+1$. The
vector space spanned by such configurations is {\footnotesize
$\left(\begin{array}{c} {\cal N}\\ p\end{array}\right)$} and the spatial hopping
term has non trivial matrix element at second order in perturbation theory. The
case $p=1$ is extensively studied in the main text sect. \ref{model}.

As also discussed in the main text, sect. \ref{sect:sun}, the SU$({\cal N})$ symmetric model $u=1$ has a very rich degeneracy any
partition of $\nu$ in ${\cal N}$ has minimal energy. This makes a perturbative
treatment in $J$ and $I$ difficult. However, it should be noted that in this
case the hopping term between the species can be treaded exactly. Indeed, by a
rotation of the Fock operators $a^\sigma$, which by definition is conserving the
total number of particles $\nu$, such term can be transformed in a species'
dependent chemical potential $m^\sigma=m+ I \lambda^\sigma$, where the
$\lambda^\sigma = 2 \cos (2\pi (\sigma-1)/{\cal N})$ are the eigenvalues of the
circulant matrix. This  means that the final ground states is non degenerate if
$I>0$ or ${\cal N}$ is even. For odd ${\cal N}=2 r+1$,  as the minimum
eigenvalue of the circulant matrix is double degenerated, $\cos(2\pi r/{\cal
N})=\cos(2\pi (r+1)/{\cal N})=-\cos(\pi/{\cal N})$, the vector space of
degenerate minima is $\nu+1$ dimensional, and the problem becomes isomorphous to
SU$(2)$ with the 
same chemical potential for the two species.    


\subsection{$p=1$ and second-order perturbation theory: the excited states and
their gaps}

Let us detail the perturbation theory in the case $0<u<1$ and $p=1$. We start by
analyzing  the effect of the hopping between different species. As its effect on
a state is to move particle to the species nearby, this term has a non zero
matrix element already at first order between two $\ket{\sigma}$ and
$\ket{\sigma^\prime}$ defined in section \ref{model}, corresponding  to the
circulant matrix
$(ZZ^\dagger)_{\sigma\sigma^\prime}=\delta_{\sigma(\sigma^\prime-1)}+\delta_{
\sigma(\sigma^\prime+1)}$, where ${\cal N}+1$ is identified with 1.         
On the contrary, the spatial hopping term has zero matrix elements between such
states as it is not conserving the on-site particle number $\nu$. This means
that we have to consider second order processes. All the possible excited states
that enter in a process constitute a vector space. A basis for them can be
obtained by applying the hopping to just one link on the link of a  product
states of $\ket{\sigma_i}$. Hence the non trivial part to be computed is of the
form $a^{\dagger\sigma}_i a^{\sigma}_{il} \ket{\sigma_i}\ket{\sigma_{il}}$. From
\begin{align}
& a^{\dagger\sigma}_i \ket{\sigma_i}= \left((1-\delta^{\sigma\sigma_i}
\sqrt{q+1} \ket{\sigma_i,\sigma} + \delta^{\sigma_i\sigma} \sqrt{q+2}
\ket{(\sigma)^2}\right),\nonumber \\
& a^{\sigma}_{il} \ket{\sigma_{il}}= \left((1-\delta^{\sigma\sigma_{il}}
\sqrt{q} \ket{\sigma_{il},(\sigma)^{-1}} + \delta^{\sigma_{il}\sigma} \sqrt{q+1}
\ket{q}\right),\label{ex}
\end{align}  
where the compact notation $\ket{(\sigma_1)^{p_1},\dots,(\sigma_r)^{p_r}}$
indicates the normalized state with $q+p_1$ particles in species $\sigma_1$,
$\dots$, $q+p_r$ particles in species $\sigma_r$, and $q$ particles in the
remaining species, it is immediate to realize that there are 4 types of excited
states of different energies. 
Indeed, the four types of a local states appearing in Eq. \ref{ex},
$\ket{\sigma,\rho}$ and $\ket{\sigma,(\rho)^{-1}}$, for any $\sigma\neq\rho$,
$\ket{(\sigma)^2}$ for any $\sigma$, and $\ket{q}$, for short $(++)$ and $(+-)$,
$(2+)$ and $(q)$ respectively. Their energy gap with respect the minimal energy
configuration are 
\begin{align}
&\Delta_{++}= -m + \frac 12 + q + \left(1 + 2 q ({\cal N} - 1)
u\right),\nonumber \\
&\Delta_{-+}=  m + \frac 12 - q - \left(1 + q ({\cal N} - 1)\right)
u,\nonumber \\
&\Delta_{2+}= -m + q + \frac 32 + q({\cal N} - 1)u,  \nonumber\\
&-\Delta_1   =  m + q + \frac 12 + q({\cal N} - 1)u.
\end{align} 
Hence, the corresponding four combination allowed for the excited states are
\begin{align}
&\Delta_{++-+}= \Delta_{++} + \Delta_{-+}=1,\nonumber\\
&\Delta_{++q} = \Delta_{++} - \Delta_1   =u,\nonumber\\
&\Delta_{2+-+}= \Delta_{2+} + \Delta_{-+}=2-u,  \nonumber\\
&\Delta_{2+q} = \Delta_{2+} - \Delta_1   =1.
\end{align}


\end{document}